\documentclass[amsmath,trackchanges]{aastex702}

\usepackage{graphicx}
\usepackage{bm}
\usepackage{booktabs}
\usepackage{threeparttable}
\usepackage{multirow}

\graphicspath{{./}{figures/}}

\shorttitle{Lunar Occultation of Planets}
\shortauthors{Guo et al.}

\begin{document}

\title{New Historical Guest-Star Candidates from Reassessed Chinese Records: Lunar Occultation of Planets}

\author[sname=Guo,gname=Jia-hao]{Jia-hao Guo}
\affiliation{Guizhou Provincial Laboratory of Big Data, State Key Laboratory of Public Big Data, Guizhou University, Guiyang, Guizhou 550025, China}
\email{gs.guojh24@gzu.edu.cn}

\author[sname=Jing,gname=Wenhui]{Wenhui Jing}
\affiliation{School of Physics and Astronomy, Yunnan University, Kunming, 650091, China}
\email{wh.jing@outlook.com}

\author[orcid=0000-0002-4421-7282,sname=Liu,gname=Yu]{Yu Liu}
\affiliation{Guizhou Provincial Laboratory of Big Data, State Key Laboratory of Public Big Data, Guizhou University, Guiyang, Guizhou 550025, China}
\email[show]{yuliu@gzu.edu.cn}

\author[sname=Sun,gname=Xiaohui]{Xiaohui Sun}
\affiliation{School of Physics and Astronomy, Yunnan University, Kunming, 650091, China}
\email{xhsun@ynu.edu.cn}

\author[sname=Pei,gname=Songpeng]{Songpeng Pei}
\affiliation{School of Physics and Electrical Engineering, Liupanshui Normal University, Liupanshui, Guizhou 553004, China}
\email{songpengpei@outlook.com}

\author[sname=Zou,gname=Yuan-Chuan]{Yuan-Chuan Zou}
\affiliation{Department of Astronomy, School of Physics, Huazhong University of Science and Technology, Wuhan 430074, China}
\email{zouyc@hust.edu.cn}

\correspondingauthor{Yu Liu}

\begin{abstract}
Ancient Chinese records of lunar occultations of planets have been used to test planetary theories and investigate long-term changes in Earth's rotation.
However, certain records contain identification errors whose origins remain poorly understood; we propose that some of these errors may arise from misidentifying guest stars, such as novae or supernovae, as planets.
We analyzed 67 errata on lunar occultations of planets from Liu Ci-Yuan's (2015) monograph using modern ephemerides and the open-source Python package \texttt{occultsky}, publicly available on GitHub.
Using recorded dates and dynastic capital coordinates, we constructed observing windows anchored at solar noon and extending by $\pm 1$ day.
Each window covered the preceding and recorded nights because the date may be off by one day.
For each of the 67 records, we cataloged the apparent right ascension and declination $(\alpha, \delta)$ at the midpoint of the Moon's visibility window on each candidate night.
The band swept by the lunar disk confines a possible transient to a narrow region a few degrees across; these midpoint coordinates serve as proxies for the implied guest-star positions.
The catalog supplies candidate positions rather than identifications, enabling remnant and compact-object searches to test the guest-star hypothesis directly.
If associations are established, these records could become dated guest-star candidates, and their dates could anchor remnant ages.
\end{abstract}

\keywords{\uat{History of astronomy}{1868} --- \uat{Novae}{1127} --- \uat{Neutron stars}{1108}}

\section{Introduction}
\label{sec:intro}

China possesses the longest continuous historical records of celestial phenomena in the world, with astronomical observations documented in official dynastic histories spanning over two millennia~\citep{1966Sci...154..597X,2002HiA....12..317S}. 
These historical records encompass all types of naked-eye astronomical phenomena, including solar and lunar eclipses, planetary motions, comets, novae and supernovae, meteors, sunspots, and aurorae.
This extraordinary observational heritage provides invaluable data for modern astronomical research, bridging the millennial-scale gap that modern instrumental observations cannot cover. 
For example, ancient planetary position records have been used to verify the accuracy of modern planetary theories~\citep{1988AJ.....96.1482H}, while historical eclipse observations spanning over 2700 years have revealed long-term fluctuations in Earth's rotation rate~\citep{1995RSPTA.351..165S}. 
Historical guest star records have played a pivotal role in identifying supernova remnants, most notably the Crab Nebula as the remnant of the 1054 CE supernova, a finding that established the foundation for historical supernova research~\citep{2008ARA&A..46..127H}.
Chinese historical nova catalogues have further systematized these ancient observations~\citep{1955AcASn...3..183X}. 
Far Eastern comet observations have provided crucial constraints on the orbital evolution of Halley's comet over two millennia~\citep{1985JBIS...38..195S}, while ancient sunspot and aurora records have revealed intense geomagnetic storms during the Medieval Maximum~\citep{2001AnGeo..19..289W}. 
These records serve as a precious bridge connecting ancient and modern astronomy.

Among these historical phenomena, lunar occultation (i.e., the Moon passing in front of a celestial object and temporarily obscuring it) holds a distinguished place in astronomical observation~\citep{1988AJ.....96.1482H}. 
This phenomenon was systematically documented in Chinese dynastic histories; for instance, the Ming Shilu alone contains 2,622 records of the Moon occulting or approaching stars~\citep{2018AcASn..59...28L}.
In modern times, advances in high-resolution ephemerides and numerical dynamical methods have revolutionized the study of lunar occultations.
By computing planetary and lunar ephemerides backward in time, modern astronomers can reconstruct the ancient sky and verify these historical records against theoretical predictions.
Such retrospective analyses not only provide critical benchmarks for calibrating ancient calendars but also furnish valuable temporal anchors for studying long-term orbital evolution in modern celestial mechanics.

Building on these methodological advances, several systematic studies have verified ancient Chinese lunar occultation records against modern ephemerides. 
\citet{1988AcApS...8..147L} pioneered the ``Really Visible Time'' method to analyze 171 Chinese lunar occultation records before AD 600, selecting 58 planetary and 113 stellar occultations to determine Earth's rotation parameters. 
\citet{1988AJ.....96.1482H} analyzed 173 historical Chinese records of occultations and appulses obtained between 146 BC and 1761 AD, using the VSOP82 planetary theory and JPL ephemeris DE 102 to verify that all well-recorded events occurred usually on the recorded day. 
\citet{1992AJ....104.2250H} focused specifically on 58 naked-eye observations of lunar occultations of planets from 68 BC to 575 AD, developing methods to determine the acceleration parameter C for Earth's rotation studies. 
\citet{1993ChA&A..17..337W} extended this approach by developing the ``time window'' method, conducting a comprehensive statistical analysis of 732 records of lunar occultations and close approaches before the Five Dynasties and filling a critical data gap from the 3rd to 7th centuries.
These studies collectively established the reliability of ancient Chinese lunar occultation records for astronomical research, while also revealing that individual records may contain errors whose origins remain poorly understood.

Despite the overall reliability of Chinese astronomical records, identification errors do occur.
For instance, the Ming Shilu alone contains 2,622 records of lunar or planetary approaches/occultations, with an error rate of 4.1\% as verified by modern astronomical computation~\citep{2018AcASn..59...28L}.
These errors fall into several categories: 
(1) misclassification of celestial objects, such as assigning the same observed phenomenon to different categories across various records~\citep{1962VA......5..127H}; 
(2) dating errors, especially for post-midnight observations that typically used the previous day's date~\citep{1981ChA&A...5..111K}; 
(3) compilation and transcription errors, since the histories were not original observations but later compilations by imperially sanctioned historians, creating the potential for copying mistakes~\citep{1988AJ.....96.1482H}; 
(4) star identification and nomenclature errors due to differing constellation naming conventions~\citep{2018AcASn..59...28L}; 
and (5) observational errors, stemming primarily from weather‑related seeing and ocular irradiation, may have led observers to record an occultation even when the objects were actually separated by several minutes of time~\citep{1988AJ.....96.1482H}.
Collectively, these studies demonstrate that while the records are generally reliable, individual errors persist and the specific mechanisms underlying the observed errors remain poorly understood. 

Among these identification errors, the confusion between guest stars and planets is particularly relevant to our study.
A compelling example is provided by the AD 1163 event: Korean historical records document a guest star on 1163 August 10, which has been proposed as a possible kilonova associated with the kilonova remnant candidate G4.8+6.2~\citep{2019MNRAS.490L..21L}. 
However, Chinese records from the same period describe a lunar encroachment (fan) on Saturn, and modern ephemeris calculations confirm that such a close encounter did occur on the corresponding date. 
This suggests that the Korean observers may have mistaken Saturn for a guest star.
Similarly, a 1645 observation of a ``guest star entering Yugui'' has been reinterpreted as a planetary transit of Mars~\citep{2025A&A...704A.251N}. 
The verb ``entering'' suggests motion, which contradicts the stationary nature of a guest star. 
Moreover, the putative nova remnant AT Cnc is located outside the Yugui asterism, and a separate Korean record from 1031 employs the very same wording to depict Mars entering that asterism~\citep{2025A&A...704A.251N}.

The reverse confusion is equally plausible: a genuine guest star or transient event could have been misidentified as a planetary occultation. 
In fact, for some historical lunar occultation records that cannot be verified by ephemeris reconstruction, the recorded ``planet'' may have been a guest star or other transient object.
Besides, this bidirectional confusion is plausible because guest stars and planets can exhibit similar apparent brightness and color to the naked eye.
This mechanism underscores the importance of evaluating all recorded lunar occultation events through modern astronomical observations and ephemeris calculations.

We present a critical reassessment of ancient Chinese records of lunar occultations of planets, specifically seeking events with potential identification errors arising from misidentification of guest stars (novae or supernovae) as planets. 
Section~\ref{sec:methodology} describes the computational methods used to reconstruct the ancient sky and cross-check the records against modern ephemerides.
Section~\ref{sec:results} presents the catalog of 67 re-examined records, reporting the midpoint coordinates of the Moon's visibility windows on the two candidate nights.
We discuss the implications of these results and the questions that remain open in Section~\ref{sec:discussion}, and summarize our conclusions in Section~\ref{sec:conclusions}.

\section{Methodology}
\label{sec:methodology}

\subsection{Data and Tool Selection}
\label{sec:data_tool_selection}

We selected Liu Ci-Yuan's (2015) monograph \textit{Investigation of Celestial Phenomena Records in Various Histories} (ISBN: 9787101103731) as our primary data source. 
This monograph systematically verifies the original records and corrects their errors through modern ephemeris reconstruction to recover historical truth.
Its scope encompasses the Twenty-Four Histories and the Draft History of the Qing Dynasty, covering thousands of entries documented in the astronomical treatises (tianwen zhi, tianxiang zhi, sitian kao, and wuxing zhi) and the basic annals (benji).
Ancient Chinese celestial records follow a concise, formulaic format typically specifying the reign period, year, month, sexagenary cycle day, and the observed phenomenon. 
Thanks to such precise dating, most records, including solar and lunar eclipses, lunar occultations, and planetary appulses to fixed stars, are verifiable in this way, while other categories such as meteors, guest stars, solar and lunar halos, and sunspots remain unverifiable.
Each entry in the monograph presents the original record text, followed by the verification analysis and textual criticism, concluding with a brief assessment, thus yielding a well-curated dataset ready for systematic use.

We extracted from the main text of the monograph all 130 entries concerning lunar occultations of planets.
From these, we excluded entries whose recorded sexagenary day does not occur in the recorded month, as well as entries lacking the year, month, or day.
These entries have incomplete or internally inconsistent dates, so they cannot be assigned reliable standard calendar dates or verified against modern ephemerides.
Some entries in the extraction are parallel accounts of the same event in different histories, so they do not represent separate events.
To avoid counting the same event more than once, only the version in the history with the earliest compilation date is retained for each group of parallel accounts.
The relevant order is \textit{Song shu} (488) $<$ \textit{Wei shu} (554) $<$ \textit{Jin shu} (648); \textit{Jiu Tang shu} (945) $<$ \textit{Xin Tang shu} (1060); \textit{Wenxian tongkao} (1307) $<$ \textit{Song shi} (1345).
Later compilers could repeat an earlier error or introduce new errors into dates, object names, or wording, so counting each version separately would inflate the catalog without adding independent evidence.
Other entries were excluded because their corrections do not concern lunar occultations of planets, including emendations of other sub-clauses within the same entry.
After these exclusions, 67 entries remain for the analysis in Section~\ref{sec:results}.
For each entry, we test the recorded occultation against modern ephemerides.
If it cannot be confirmed, we consider whether a guest star or other transient object may have been misidentified as a planet and use the reconstructed lunar trajectory to constrain the transient's possible position.

We selected the Skyfield Python library for our analysis because of its high-precision ephemeris calculations and flexible Python interface~\citep{2019ascl.soft07024R}. 
Skyfield relies on planetary ephemerides developed by the Jet Propulsion Laboratory (JPL) to compute the state vectors of solar system bodies. 
By loading binary ephemeris files and applying interpolation algorithms, it determines object positions and velocities at any given time.
Furthermore, the library not only supports transformations between multiple astronomical reference frames, including geocentric equatorial, ecliptic, and horizontal coordinates, but also precisely computes lunar phases, conjunctions between the Moon and planets, and sunrise and sunset times.
These capabilities enable researchers to determine whether the observational conditions at the time permitted the naked-eye visibility of recorded phenomena, making Skyfield an ideal tool for verifying historical records of guest stars, comets, and other celestial events.

JPL provides several planetary ephemerides, among which DE440 is the most widely used for modern astronomical calculations, covering the period 1849--2150.\footnote{The ephemeris files can be downloaded from \url{https://naif.jpl.nasa.gov/pub/naif/generic_kernels/spk/planets/}} 
However, DE440 does not span the full temporal range of our dataset (303--1741 CE). 
To address this limitation, we selected DE441, which includes two extended components: \texttt{de441\_part-1.bsp} (13201 BCE--1969 CE) and \texttt{de441\_part-2.bsp} (1969 CE--17191 CE). 
Together these files provide continuous coverage exceeding 30,000 years, making DE441 the optimal choice for ancient astronomical calculations. 

Because the Skyfield library defaults to the proleptic Gregorian calendar, special care is required when processing historical Chinese dates. 
The traditional Chinese calendar system centers on dynasty, era name, and the sexagenary cycle day, forming a unique chronological expression method.
To ensure precise backtracking and verification of celestial positions in modern scientific analysis, these traditional dating records must be converted to standard calendar formats.
Specifically, dates before October 15, 1582 (the effective date of the Gregorian calendar reform) should use the Julian calendar. 
Skyfield controls this transition point with the `julian\_calendar\_cutoff` parameter.
Traditionally, such conversions rely on manual consultation of specialized chronological tables, such as the \textit{Zhongguo shi li ri he Zhong xi li ri dui zhao biao} (ISBN: 9787208070165).

To automate this conversion process, we developed a Python script that interfaces programmatically with the Two Thousand Years of Chinese and Western Calendar Conversion tool maintained by the Academia Sinica Center for Digital Cultures\footnote{\url{https://sinocal.sinica.edu.tw/}}.
The web interface presents a three-level cascading form (dynasty, emperor, and reign period) whose option lists are dynamically populated by JavaScript upon each selection.
We therefore simulated browser interactions to traverse all three dropdown levels and extract the complete mapping from dynasty through emperor to reign period, thereby overcoming the JavaScript-driven dynamic content challenge.
A further complication arises from the fact that historical records employ free-text names that frequently include variant or abbreviated forms of dynastic and imperial titles, whereas the conversion system requires exact Traditional Chinese character strings, necessitating a normalization step.
To address this, we constructed a normalization table that maps each textual designation in our dataset to its canonical form accepted by the system.
Additionally, records containing intercalary lunar months required an additional flag to distinguish them from ordinary months.
Finally, for each record, the calibrated dynasty, emperor, and reign-period values, together with the year, month, day, and any intercalary-month flag, were submitted to the conversion endpoint.
The script then parsed the returned HTML to extract the corresponding Gregorian date.

To compute the apparent positions of the Moon and planets as observed from ancient China, we must specify the geographic coordinates of the observation site. 
Chinese astronomical observations were typically conducted at the imperial capital of each dynasty, which changed location over the two-millennium span of our records~\citep{1988AJ.....96.1482H,2020JKAS...53..125A}. 
We compiled a lookup table mapping each dynasty to its capital city's geographic coordinates, following the methodology established by Hilton et al.~\citep{1988AJ.....96.1482H,1992AJ....104.2250H}. 
For dynasties with multiple administrative centers, we adopted the primary capital as the representative observation site; for example, the Cao Wei period (220–265 CE) is represented by Luoyang, with its secondary capitals Xuchang and Ye (modern Linzhang) consolidated into this entry.
Table~\ref{tab:capitals} lists the geographic coordinates of capital cities corresponding to dynasties represented in our 67 validated records. 
The table includes only dynasties that have surviving records in our dataset, along with the specific capital cities used for ephemeris calculations. 
For dynasties that relocated their capitals, the table notes the capital transitions with corresponding date ranges, ensuring that each record is assigned the appropriate observational coordinates based on its historical period. 
The geographic coordinates (longitude and latitude) are sourced from the China Historical Geographic Information System (CHGIS) Time-Space Gazetteer (TGAZ), a joint project of Harvard University and Fudan University, which provides historically accurate locations rather than modern city centers. 
The altitude values were sampled at the same coordinates from the Copernicus Digital Elevation Model GLO-90 (ESA, 2020, 90 m resolution).

\begin{table}[htbp]
\centering
\begin{threeparttable}
\caption{Geographic coordinates of capital cities used for ephemeris calculations, showing only dynasties represented in the 67 validated records.}
\label{tab:capitals}

\begin{tabular}{llccc}
\toprule
City & Dynasty & Longitude (°E) & Latitude (°N) & Altitude (m) \\
\midrule
Xi'an     & Western Jin\tnote{a}, Tang           & 108.91  & 34.25  & 411   \\
Luoyang   & Cao Wei, Western Jin\tnote{a}, Northern Wei\tnote{b}, Later Tang & 112.60  & 34.73  & 125   \\
Nanjing   & Eastern Jin, Liu Song, Southern Qi, Ming\tnote{d} & 118.77  & 32.05  & 19    \\
Datong    & Northern Wei\tnote{b}       & 113.30  & 40.09  & 1053  \\
Kaifeng   & Later Jin, Northern Song, Jin\tnote{c} & 114.34  & 34.79  & 75    \\
Hangzhou  & Southern Song               & 120.17  & 30.29  & 10    \\
Beijing   & Jin\tnote{c}, Yuan, Ming\tnote{d}, Qing & 116.37  & 39.93  & 52  \\
Linzhang  & Northern Qi, Eastern Wei    & 114.41  & 36.27  & 78    \\
\bottomrule
\end{tabular}

\begin{tablenotes}
\footnotesize
\item[a] Western Jin: Capital at Luoyang (265--313), then moved to Xi'an (313--316).
\item[b] Northern Wei: The earliest capital (Shengle) is excluded from this study. Capital at Datong (398--494), then moved to Luoyang (494--534). 
\item[c] Jin: The earliest capital (Huining) is excluded from this study. Capital at Beijing (1153--1214), then moved to Kaifeng (1214--1234). 
\item[d] Ming: Capital at Nanjing (1368--1420), then moved to Beijing (1420--1644).
\end{tablenotes}

\end{threeparttable}
\end{table}

Historical astronomical records frequently reference stellar positions relative to specific stars or constellations, such as the Twenty-Eight Mansions (Er Shi Ba Xiu) of ancient China. 
To reconstruct these celestial configurations within their proper cultural and historical context, we employ Stellarium, an open-source planetarium software that supports multiple sky cultures~\citep{2021arXiv210401019Z}. 
Stellarium's Sky Cultures module provides a diverse collection of constellation traditions from various civilizations, including Chinese, Arabic, Greek, and indigenous systems~\citep{2026AAS...24712503M}. 
This capability is essential for cultural astronomy research, as it enables the reconstruction of the night sky from the perspective of ancient observers, thereby facilitating a deeper understanding of historical astronomical records. 
For instance, when simulating Chinese guest star records, the software utilizes the traditional Chinese star mansion system rather than Western constellation divisions. 
The complete collection of sky cultures, which typically includes constellation lines, illustrations, common names, and cultural descriptions, is maintained in the Stellarium Sky Cultures repository\footnote{\url{https://github.com/guillaumeduprat/stellarium-skycultures}}.

\subsection{Computational Methods}
\label{sec:computational_methods}

With the dataset, ephemeris, and observer coordinates established above, several computational details remain to be addressed before the historical records can be verified against modern predictions. 
These center on the observational window, defined from astronomical twilight times together with lunar and planetary rise and set times, and on the calculation of the angular separation between the Moon and the target planet within this window to identify occultation events. 
Additional details include the conversion between local and Universal Time and the evaluation of the Moon's and planets' visibility and apparent positions. 
We implemented these computations as methods of an \texttt{Observer} class, the core of our Python-based analysis framework.\footnote{\url{https://doi.org/10.5281/zenodo.24297100}}

A significant complication arises from the Chinese chronological convention of recording nocturnal observations. 
In traditional Chinese annals, the calendar day begins at midnight, yet observations made during the latter half of the night (after midnight) were frequently recorded under the date of the preceding evening~\citep{1981ChA&A...5..111K}. 
Consequently, an event recorded under a given date may have occurred in the early morning hours (00:00--08:00 local time) of that date, which belong to the preceding night, or in the early morning hours of the following day, which belong to the night of the recorded date. 
To account for this ambiguity, we anchor the observational window at the solar noon of the recorded date and extend it by ±1 day, thereby covering both the night preceding the recorded date and the night of the recorded date itself. 
This approach directly affects the determination of the observational window for each historical record.

The ancient Chinese timekeeping system, calibrated by sundials and sustained by water clocks, approximates Local Mean Time (LMT). 
To interface with the Skyfield library, which requires Universal Time (UTC) as input, we convert local times using a two-step transformation. 
First, we apply a longitude-based offset to derive LMT: UTC = LMT - (longitude / 15°) hours, with the capital longitudes provided in Table~\ref{tab:capitals}; for example, observations made in Hangzhou (120.17°E) require a correction of approximately -8 hours (UTC+08:00) relative to Greenwich. 
Second, we compute solar noon, the instant when the Sun reaches its highest altitude, using Skyfield's almanac transit-finding algorithm to determine the precise UTC time of the Sun's upper meridian transit, thereby establishing Local Apparent Time (LAT), which aligns with the ancient Chinese sundial-based system. 
The difference between LAT and LMT, known as the Equation of Time, reaches a maximum of approximately ±16 minutes. 
Since this discrepancy is small, and because a timezone offset provides a convenient conversion between LMT and UTC, our computational framework outputs results primarily in LMT, with solar noon serving as a reference point. 
This approach is sufficient for our objective of determining the Moon's visibility window rather than pinpointing the exact ancient moment.

To determine the Moon's visibility window on each night, we compute the intersection of two conditions: darkness and lunar visibility. 
First, we identify periods of darkness using astronomical twilight times. 
These mark when the Sun descends more than 12° below the horizon, the threshold for naked-eye astronomical observations. 
Second, we calculate rise and set times for both the Moon and each target planet using Skyfield's almanac functions to identify when they are above the horizon. 
The resulting rise and set times are converted to local time using the observer's timezone offset, allowing us to determine which planets are visible during this window. 
This information is crucial for evaluating whether historical observations were physically possible at the recorded times and locations. 

Additionally, we compute lunar phase and angular radius at midnight (solar noon + 12 hours) for each observation date. 
Since the lunar phase changes slowly over the course of a day, a single calculation at midnight suffices to characterize the Moon's illumination state and apparent size for the entire observational window. 
The phase is computed as the angular separation between the Sun and Moon, yielding a phase angle from 0° (new moon) to 180° (full moon). 
The angular radius is derived from the topocentric distance to the Moon's center, accounting for the Moon's physical radius and the observer's position. 
These lunar properties are critical for evaluating occultation events, as they affect both the visibility of the Moon and the geometric conditions required for a successful occultation.

The central metric for occultation identification is the angular separation between the Moon and each target planet, with a threshold of 2° adopted consistent with Liu's methodology~\citep{1988AJ.....96.1482H}. 
For each time sample within the observational window, we compute the topocentric apparent positions of both the Moon and the target planet using Skyfield's vector-based calculations. 
We then derive their angular separation in degrees. 
We retain the minimum Moon--planet separation across all time samples as the predicted occultation time. 
This approach enables the systematic identification of occultation events across the entire dataset. 
When a recorded planetary occultation cannot be confirmed against modern ephemerides, the alternative is that a guest star erupted that night along the Moon's path. 
The Moon occulted it and drew the observer's attention, who recorded it as a planetary occultation. 
Because the Moon shifts roughly 13° per day against the stellar background, the same alignment would not recur on the following night, and the event would be recorded only once. 
The guest star, though potentially still visible, would escape further notice.

To reconstruct the possible position of such a guest star, we compute the Moon's trajectory across the sky on the night of the recorded event.
For each time sample within the observational window, the framework computes the Moon's right ascension and declination, tracing the region swept by the lunar disk. 
A guest star falling within this region would have been occulted by the Moon and recorded as a planet.
We project this region onto a sky map using the STG (Stereographic) projection, a conformal map projection widely used in the FITS World Coordinate System (WCS) standard~\citep{2002A&A...395.1077C}.
The stereographic projection preserves local angles, ensuring that the shapes of constellations and asterisms are accurately represented with minimal distortion at the edges~\citep{2022MNRAS.509.4480W}.
For the moderate-sized sky regions typical of Chinese star mansions (er shi ba xiu), the STG projection provides a natural and realistic visual representation that closely approximates the appearance of the night sky as observed by the naked eye in ancient China.
The sky map also labels prominent stars using the traditional Chinese star mansion names from Stellarium's Sky Cultures module, enabling direct cross-referencing with the terminology in the historical records.

\subsection{Integrated Workflow}
\label{sec:integrated_workflow}

The complete workflow for searching and verifying historical occultation events proceeds through four integrated stages. 
First, for each historical record, we retrieve the validated Western calendar date obtained from the conversion described in Section~\ref{sec:data_tool_selection}, identify the dynasty, and determine the capital's geographic coordinates (Table~\ref{tab:capitals}). 
Second, we construct an Observer instance at the dynastic capital and compute the observational window anchored at solar noon and extended by ±1 day, along with the twilight times, lunar and planetary rise and set times, and the lunar phase and angular radius described in Section~\ref{sec:computational_methods}. 
Third, we sample the observational window at 15-minute intervals, evaluate the Moon--planet angular separation at each epoch, and identify occultation events when the minimum separation falls below the 2° threshold~\citep{1988AJ.....96.1482H}. 
Fourth, for events where the recorded occultation cannot be confirmed against modern ephemerides, we reconstruct the Moon's trajectory and project it onto a sky map labeled with Chinese star mansion and star names, using the STG projection described in Section~\ref{sec:computational_methods}. 
This sky map pinpoints the position of a candidate guest star that may have been misidentified as a planet.

To validate the reliability of our integrated workflow, we present a detailed case study of the AD 1163 event, which has been proposed as a possible kilonova associated with the remnant candidate G4.8+6.2~\citep{2019MNRAS.490L..21L}. 
As discussed in Section~\ref{sec:intro}, Chinese records document a lunar encroachment on Saturn on 1163 August 10. 
Because the record of this event is well established, it provides an ideal benchmark for testing whether our workflow reproduces the recorded event.

\begin{figure}[htbp]
    \centering
    \includegraphics[width=0.95\textwidth]{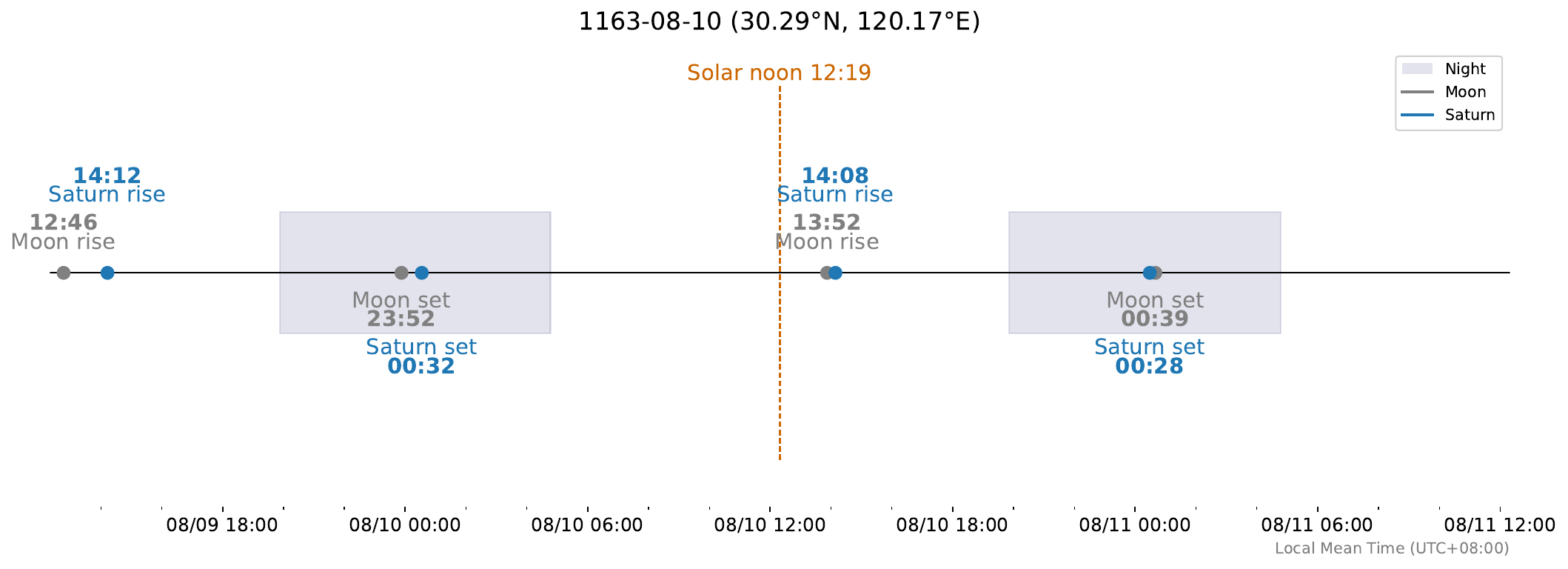}
    \caption{Timeline visualization of the AD 1163 lunar encroachment on Saturn.}
    \label{fig:timeline_1163}
\end{figure}

Figure~\ref{fig:timeline_1163} presents a timeline visualization of the AD 1163 event, showing the key astronomical events for the night of 1163 August 10 as observed from the Southern Song capital at Hangzhou (120.17°E, 30.29°N). 
The timeline displays the local mean time (LMT) axis spanning solar noon ±1 day, with night periods indicated by semi-transparent bands around the time axis. 
Rise and set events for the Moon and Saturn are marked as colored dots with time and body labels, with rise events above the axis and set events below. 
The vertical dashed line marks solar noon, and the x-axis label indicates the local mean time offset (UTC+08:00).
The timeline provides a comprehensive overview of the observational window, illustrating that both the Moon and Saturn were visible on the nights of August 9 and 10. 
Notably, the two bodies set at nearly the same time at the end of the night of August 10. 

\begin{figure}[htbp]
    \centering
    \includegraphics[width=0.95\textwidth]{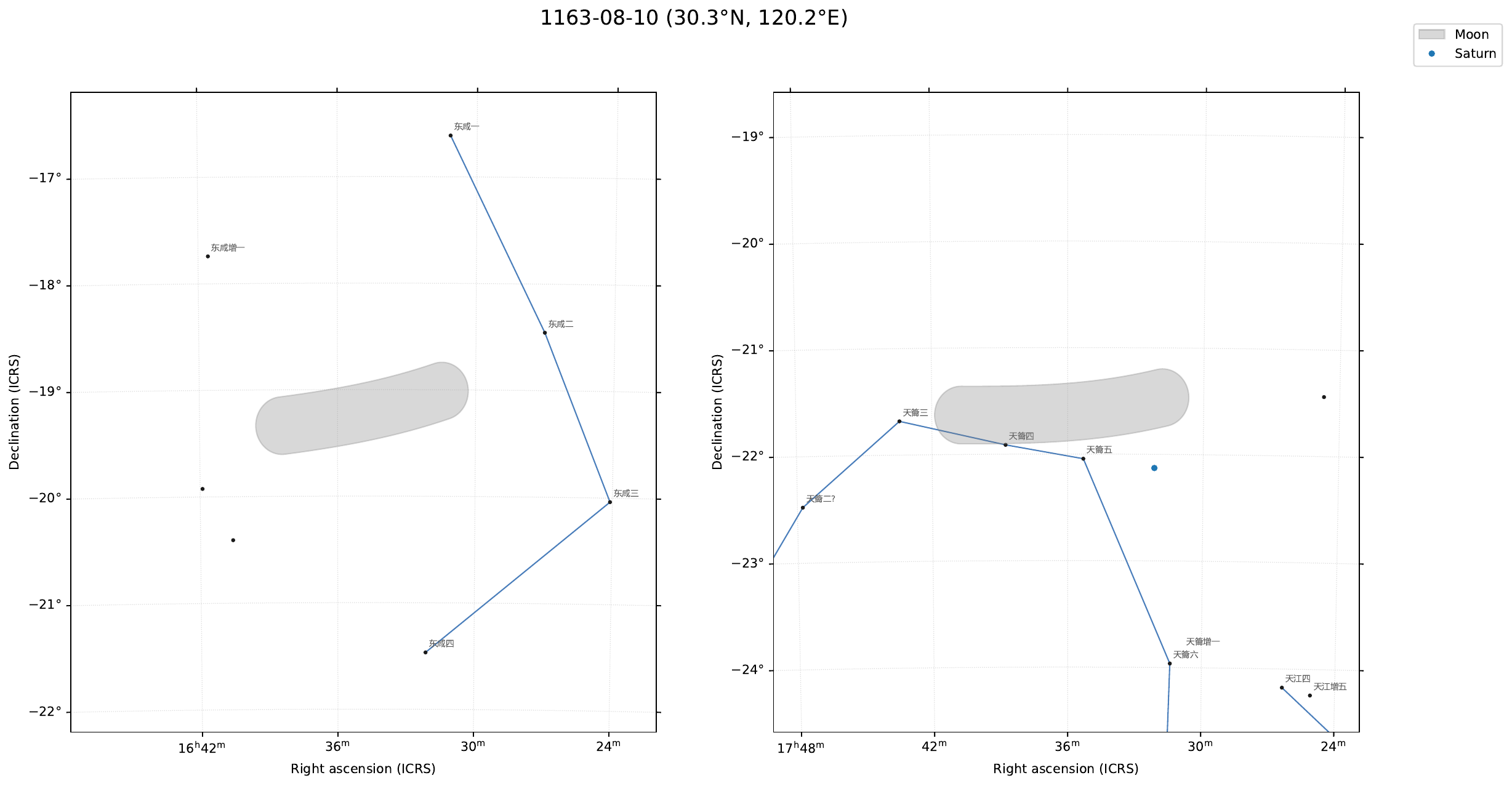}
    \caption{Sky map visualization of the AD 1163 lunar encroachment on Saturn.}
    \label{fig:skyview_1163}
\end{figure}

Figure~\ref{fig:skyview_1163} presents the trajectory band swept by the lunar disk over the two-day observational window determined from the timeline. 
Saturn moved less than 0.01° during the night, so its trace is essentially a single point.
The map is also labeled with the traditional Chinese star-mansion and star names from Stellarium’s Sky Cultures module (Section~\ref{sec:computational_methods}). 
Over the course of the night, the Moon advanced eastward, approaching Saturn and reaching minimum separation at 20:23 LMT.
At this minimum center-to-center separation of 0.631°, the lunar limb remained well clear of Saturn’s ring system and planetary limb, consistent with the event being recorded as a lunar encroachment.
At that moment, the Moon and Saturn lay close to the fifth star of the Tianyue asterism. 
Later that night, the Moon approached the fourth star of the Tianyue asterism. 
This reconstruction thus confirms that our workflow reproduces the recorded event, passing the AD 1163 benchmark. 

\section{Results}
\label{sec:results}

The catalog comprises the 67 errata on lunar occultations of planets in Liu Ci-Yuan's (2015) monograph that remain after the selection described in Section~\ref{sec:data_tool_selection}, which excludes entries lacking convertible dates, duplicates, and corrections unrelated to such occultations.
The entries span 303--1741 CE and draw on 11 histories, covering 14 dynasties and eight observing capitals.
Using the computational methods of Section~\ref{sec:computational_methods}, implemented in our Python package \texttt{occultsky}, we computed the midpoint coordinates of the Moon's visibility windows on the two candidate nights for each entry, and the results are collected in Table~\ref{tab:events}.
Each midpoint stands for the trajectory band swept by the lunar disk on that night.
The band is plotted in full as a sky map, as illustrated for the AD 1163 benchmark in Figure~\ref{fig:skyview_1163}.
A transient misidentified as the recorded planet must have lain within this band, so the midpoints serve as proxies for the implied guest-star positions.

\begingroup
\scriptsize
\setlength{\LTleft}{0pt}
\setlength{\tabcolsep}{4pt}
\begin{longtable}{>{\raggedright\arraybackslash}p{\dimexpr0.29\linewidth} >{\raggedright\arraybackslash}p{\dimexpr0.14\linewidth} >{\raggedright\arraybackslash}p{\dimexpr0.095\linewidth} >{\raggedright\arraybackslash}p{\dimexpr0.18\linewidth} >{\raggedright\arraybackslash}p{\dimexpr0.18\linewidth}}
\caption{Catalog of the 67 errata entries.\label{tab:events}}\\
\toprule
Book \& Section & Dynasty & Date & Moon (ra, dec) & Moon (ra, dec) \\
(Entry No.) & Capital & & (preceding night) & (recorded night) \\
\midrule
\endfirsthead
\caption[]{Catalog of the 67 errata entries (continued).}\\
\toprule
Book \& Section & Dynasty & Date & Moon (ra, dec) & Moon (ra, dec) \\
(Entry No.) & Capital & & (preceding night) & (recorded night) \\
\midrule
\endhead
\midrule
\multicolumn{5}{r}{\footnotesize Continued on next page} \\
\endfoot
\bottomrule
\endlastfoot
{}[\textit{Jin shu}, Yue yan fan wu xing] entry 1924 & Western~Jin, \mbox{Luoyang} & 303-12-05 & (22.24, 8.09) & (34.05, 11.42) \\
{}[\textit{Jin shu}, Yue yan fan wu xing] entry 1944 & Eastern~Jin, \mbox{Nanjing} & 338-05-24 & (307.49, -15.26) & (319.75, -11.32) \\
{}[\textit{Jin shu}, Yue yan fan wu xing] entry 1944 & Eastern~Jin, \mbox{Nanjing} & 338-08-28 & (122.78, 15.26) & (137.20, 11.07) \\
{}[\textit{Song shu}, Tianwen zhi II, ch.~24] entry 2709 & Eastern~Jin, \mbox{Nanjing} & 322-01-06 & (322.39, -10.76) & (335.50, -6.73) \\
{}[\textit{Song shu}, Tianwen zhi II, ch.~24] entry 2886 & Eastern~Jin, \mbox{Nanjing} & 359-04-20 & (114.95, 15.97) & (127.22, 13.57) \\
{}[\textit{Song shu}, Tianwen zhi III, ch.~25] entry 2967 & Eastern~Jin, \mbox{Nanjing} & 400-07-23 & (329.70, -15.41) & (341.69, -11.87) \\
{}[\textit{Song shu}, Tianwen zhi III, ch.~25] entry 3007 & Eastern~Jin, \mbox{Nanjing} & 405-06-11 & (74.20, 25.78) & (90.10, 27.33) \\
{}[\textit{Song shu}, Tianwen zhi IV, ch.~26] entry 3187 & Liu~Song, \mbox{Nanjing} & 434-05-09 & (241.95, -17.63) & (255.03, -18.91) \\
{}[\textit{Song shu}, Tianwen zhi IV, ch.~26] entry 3187 & Liu~Song, \mbox{Nanjing} & 434-05-20 & (18.19, 6.01) & (31.31, 9.85) \\
{}[\textit{Song shu}, Tianwen zhi IV, ch.~26] entry 3328 & Liu~Song, \mbox{Nanjing} & 466-05-25 & (17.14, 12.18) & (30.76, 17.15) \\
{}[\textit{Nan Qi shu}, Yue fan lie xing] entry 3618 & Southern~Qi, \mbox{Nanjing} & 483-10-21 & (262.06, -27.86) & (277.81, -26.91) \\
{}[\textit{Nan Qi shu}, Yue fan lie xing] entry 3653 & Southern~Qi, \mbox{Nanjing} & 488-07-03 & (216.33, -15.50) & (230.30, -18.31) \\
{}[\textit{Nan Qi shu}, Yue fan lie xing] entry 3690 & Southern~Qi, \mbox{Nanjing} & 492-01-17 & --- & (347.04, -7.40) \\
{}[\textit{Nan Qi shu}, Yue fan lie xing] entry 3767 & Southern~Qi, \mbox{Nanjing} & 494-08-09 & (53.14, 14.62) & (67.70, 18.45) \\
{}[\textit{Wei shu}, Tianxiang zhi II, ch.~105.2] entry 4333 & Northern~Wei, \mbox{Datong} & 412-02-18 & (212.29, -17.37) & (226.33, -20.72) \\
{}[\textit{Wei shu}, Tianxiang zhi II, ch.~105.2] entry 4368 & Northern~Wei, \mbox{Datong} & 420-12-21 & --- & --- \\
{}[\textit{Wei shu}, Tianxiang zhi II, ch.~105.2] entry 4480 & Northern~Wei, \mbox{Datong} & 480-03-15 & (226.32, -20.91) & (241.69, -25.52) \\
{}[\textit{Wei shu}, Tianxiang zhi II, ch.~105.2] entry 4653 & Northern~Wei, \mbox{Luoyang} & 529-11-19 & (274.13, -20.73) & (287.53, -20.64) \\
{}[\textit{Wei shu}, Tianxiang zhi IV, ch.~105.4] entry 4952 & Northern~Wei, \mbox{Luoyang} & 498-03-21 & (161.68, 11.89) & (175.37, 5.00) \\
{}[\textit{Wei shu}, Tianxiang zhi IV, ch.~105.4] entry 4952 & Northern~Wei, \mbox{Luoyang} & 498-05-01 & (353.21, -7.56) & (3.83, -1.94) \\
{}[\textit{Wei shu}, Tianxiang zhi IV, ch.~105.4] entry 5046 & Eastern~Wei, \mbox{Linzhang} & 539-04-22 & (259.35, -28.15) & (273.01, -27.81) \\
{}[\textit{Jiu Tang shu}, Benji, ch.~1--20] entry 5959 & Tang, \mbox{Xi'an} & 839-08-28 & (342.52, -2.76) & (356.04, 3.11) \\
{}[\textit{Jiu Tang shu}, Benji, ch.~1--20] entry 5979 & Tang, \mbox{Xi'an} & 886-05-08 & --- & --- \\
{}[\textit{Jiu Tang shu}, Zaiyi] entry 6125 & Tang, \mbox{Xi'an} & 712-09-21 & (28.84, 10.39) & (41.77, 13.55) \\
{}[\textit{Xin Tang shu}, Yue wuxing lingfan ji xingbian] entry 6749 & Tang, \mbox{Xi'an} & 807-03-07 & (299.47, -17.08) & (311.96, -14.97) \\
{}[\textit{Xin Tang shu}, Yue wuxing lingfan ji xingbian] entry 6754 & Tang, \mbox{Xi'an} & 816-12-25 & (318.71, -18.75) & (331.11, -13.39) \\
{}[\textit{Xin Tang shu}, Yue wuxing lingfan ji xingbian] entry 6779 & Tang, \mbox{Xi'an} & 823-02-12 & (310.84, -13.74) & (323.10, -10.15) \\
{}[\textit{Xin Tang shu}, Yue wuxing lingfan ji xingbian] entry 6844 & Tang, \mbox{Xi'an} & 833-07-11 & (3.75, -0.88) & (14.43, 4.86) \\
{}[\textit{Xin Tang shu}, Yue wuxing lingfan ji xingbian] entry 6859 & Tang, \mbox{Xi'an} & 835-11-02 & (334.38, -11.75) & (346.28, -5.65) \\
{}[\textit{Xin Tang shu}, Yue wuxing lingfan ji xingbian] entry 6869 & Tang, \mbox{Xi'an} & 837-03-15 & (61.69, 25.56) & (75.99, 26.83) \\
{}[\textit{Xin Tang shu}, Yue wuxing lingfan ji xingbian] entry 6884 & Tang, \mbox{Xi'an} & 839-04-02 & (195.25, -12.40) & (206.87, -16.47) \\
{}[\textit{Xin Tang shu}, Yue wuxing lingfan ji xingbian] entry 6919 & Tang, \mbox{Xi'an} & 900-12-17 & (195.43, -3.34) & (206.96, -7.17) \\
{}[\textit{Xin Wudai shi}, Sitian kao II, ch.~59] entry 7171 & Later~Tang, \mbox{Luoyang} & 927-01-01 & (226.56, -21.22) & (239.90, -25.14) \\
{}[\textit{Xin Wudai shi}, Sitian kao II, ch.~59] entry 7226 & Later~Jin, \mbox{Kaifeng} & 944-08-01 & (239.62, -24.17) & (253.72, -27.15) \\
{}[\textit{Song shi}, Benji, ch.~1--47] entry 7339 & Northern~Song, \mbox{Kaifeng} & 1040-03-21 & (73.01, 27.59) & (86.91, 28.24) \\
{}[\textit{Song shi}, Benji, ch.~1--47] entry 7344 & Northern~Song, \mbox{Kaifeng} & 1046-04-22 & (204.76, -12.73) & (219.52, -16.43) \\
{}[\textit{Song shi}, Benji, ch.~1--47] entry 7474 & Yuan, \mbox{Beijing} & 1277-09-11 & (334.47, -16.79) & (346.68, -11.85) \\
{}[\textit{Song shi}, Yue fan wu wei] entry 7681 & Northern~Song, \mbox{Kaifeng} & 978-08-22 & (359.19, -6.58) & (12.52, -0.78) \\
{}[\textit{Song shi}, Yue fan wu wei] entry 7681 & Northern~Song, \mbox{Kaifeng} & 978-09-26 & (98.43, 23.59) & (111.73, 23.41) \\
{}[\textit{Song shi}, Yue fan wu wei] entry 7686 & Northern~Song, \mbox{Kaifeng} & 988-03-11 & (236.84, -23.14) & (251.39, -24.43) \\
{}[\textit{Song shi}, Yue fan wu wei] entry 7691 & Northern~Song, \mbox{Kaifeng} & 992-05-04 & --- & --- \\
{}[\textit{Song shi}, Yue fan wu wei] entry 7696 & Northern~Song, \mbox{Kaifeng} & 994-03-31 & (217.32, -10.53) & (232.67, -14.57) \\
{}[\textit{Song shi}, Yue fan wu wei] entry 7701 & Northern~Song, \mbox{Kaifeng} & 1004-09-07 & (60.47, 25.18) & (74.10, 26.59) \\
{}[\textit{Song shi}, Yue fan wu wei] entry 7711 & Northern~Song, \mbox{Kaifeng} & 1011-09-25 & (131.09, 14.02) & (145.19, 10.82) \\
{}[\textit{Song shi}, Yue fan wu wei] entry 7716 & Northern~Song, \mbox{Kaifeng} & 1020-03-10 & (154.21, 13.73) & (168.03, 6.85) \\
{}[\textit{Song shi}, Yue fan wu wei] entry 7721 & Northern~Song, \mbox{Kaifeng} & 1021-10-15 & (297.35, -25.51) & (312.11, -21.25) \\
{}[\textit{Song shi}, Yue fan wu wei] entry 7726 & Northern~Song, \mbox{Kaifeng} & 1038-04-19 & (180.45, 1.00) & (193.78, -6.24) \\
{}[\textit{Song shi}, Yue fan wu wei] entry 7731 & Northern~Song, \mbox{Kaifeng} & 1041-11-16 & (134.27, 16.83) & (146.12, 11.98) \\
{}[\textit{Song shi}, Yue fan wu wei] entry 7741 & Northern~Song, \mbox{Kaifeng} & 1056-01-07 & (142.55, 19.85) & (156.58, 14.44) \\
{}[\textit{Song shi}, Yue fan wu wei] entry 7741 & Northern~Song, \mbox{Kaifeng} & 1056-01-10 & (183.18, 1.32) & (196.18, -5.55) \\
{}[\textit{Song shi}, Yue fan wu wei] entry 7746 & Northern~Song, \mbox{Kaifeng} & 1056-04-01 & (178.44, 3.25) & (191.82, -3.93) \\
{}[\textit{Song shi}, Yue fan wu wei] entry 7751 & Northern~Song, \mbox{Kaifeng} & 1056-06-22 & (177.70, 2.89) & (190.09, -3.74) \\
{}[\textit{Song shi}, Yue fan wu wei] entry 7756 & Northern~Song, \mbox{Kaifeng} & 1057-07-14 & (232.29, -24.04) & (248.00, -27.62) \\
{}[\textit{Song shi}, Yue fan wu wei] entry 7771 & Southern~Song, \mbox{Hangzhou} & 1209-07-06 & --- & (152.11, 8.20) \\
{}[\textit{Song shi}, Yue fan wu wei] entry 7771 & Southern~Song, \mbox{Hangzhou} & 1209-07-25 & (36.42, 19.45) & (50.50, 23.19) \\
{}[\textit{Song shi}, Yue fan wu wei] entry 7778 & Southern~Song, \mbox{Hangzhou} & 1237-08-04 & (271.32, -21.55) & (285.44, -21.87) \\
{}[\textit{Yuan shi}, Benji, ch.~1--47] entry 10256 & Yuan, \mbox{Beijing} & 1289-08-05 & (352.23, -2.18) & (6.70, 2.66) \\
{}[\textit{Yuan shi}, Benji, ch.~1--47] entry 10408 & Yuan, \mbox{Beijing} & 1333-10-12 & --- & (245.50, -23.33) \\
{}[\textit{Yuan shi}, Benji, ch.~1--47] entry 10418 & Yuan, \mbox{Beijing} & 1334-04-07 & (55.02, 20.10) & (70.42, 24.03) \\
{}[\textit{Yuan shi}, Yue wuxing lingfan ji xingbian shang] entry 10786 & Yuan, \mbox{Beijing} & 1284-01-13 & (235.09, -25.18) & (248.84, -26.95) \\
{}[\textit{Yuan shi}, Yue wuxing lingfan ji xingbian shang] entry 10868 & Yuan, \mbox{Beijing} & 1295-02-23 & (63.57, 19.13) & (76.43, 21.82) \\
{}[\textit{Yuan shi}, Yue wuxing lingfan ji xingbian shang] entry 10945 & Yuan, \mbox{Beijing} & 1308-02-16 & (256.07, -18.92) & (269.55, -19.16) \\
{}[\textit{Ming shi}, Yue yanfan wuwei] entry 11120 & Ming, \mbox{Nanjing} & 1379-03-19 & --- & --- \\
{}[\textit{Ming shi}, Yue yanfan wuwei] entry 11125 & Ming, \mbox{Nanjing} & 1390-12-21 & (76.09, 26.36) & (92.50, 27.74) \\
{}[\textit{Ming shi}, Yue yanfan wuwei] entry 11125 & Ming, \mbox{Nanjing} & 1406-02-15 & (294.52, -23.32) & (308.73, -21.69) \\
{}[\textit{Ming shi}, Yue yanfan wuwei] entry 11132 & Ming, \mbox{Beijing} & 1445-12-20 & (170.15, 8.87) & (182.74, 2.61) \\
{}[\textit{Ming shi}, Yue yanfan wuwei] entry 11137 & Ming, \mbox{Beijing} & 1449-04-24 & (57.33, 24.67) & (73.51, 27.20) \\
{}[\textit{Ming shi}, Yue yanfan wuwei] entry 11142 & Ming, \mbox{Beijing} & 1454-03-04 & (38.46, 17.14) & (50.99, 19.57) \\
{}[\textit{Ming shi}, Yue yanfan wuwei] entry 11147 & Ming, \mbox{Beijing} & 1456-05-18 & (218.67, -14.86) & (233.47, -17.69) \\
{}[\textit{Ming shi}, Yue yanfan wuwei] entry 11157 & Ming, \mbox{Beijing} & 1478-04-08 & (90.39, 18.98) & (105.98, 19.01) \\
{}[\textit{Ming shi}, Yue yanfan wuwei] entry 11167 & Ming, \mbox{Beijing} & 1586-10-09 & (150.53, 7.15) & (164.86, 2.52) \\
{}[\textit{Ming shi}, Yue yanfan wuwei] entry 11172 & Ming, \mbox{Beijing} & 1607-06-27 & --- & (137.44, 13.40) \\
{}[\textit{Qing shi gao}, Tianwen zhi XI, ch.~36] entry 11995 & Qing, \mbox{Beijing} & 1736-05-08 & (15.21, 5.03) & --- \\
{}[\textit{Qing shi gao}, Yue wuxing lingfan yanju] entry 12246 & Qing, \mbox{Beijing} & 1741-01-14 & (250.48, -20.43) & (265.12, -22.83) \\
\bottomrule
\end{longtable}
\endgroup

In Table~\ref{tab:events}, the first column identifies each entry by its source book and section in square brackets (with the chapter number when given), followed by the entry number, and the second names the dynasty and its observing capital.
These entry numbers are the enumeration of Liu's (2015) modern compilation; the original historical texts carry no entry numbering of their own.
Dates in the third column are converted to the Gregorian calendar in the ``YYYY-MM-DD'' format. 
An entry may contain several sub-clauses that carry different dates; each date occupies one row, so the 67 entries occupy 74 rows.
The entries are ordered by book (in chronological order of the histories, from the \textit{Jin shu} to the \textit{Qing shi gao}) and by entry number within each book, rather than by date.
The last two columns give the Moon's apparent right ascension and declination in degrees, as an ordered pair $(\alpha, \delta)$ rounded to two decimal places, at the midpoint of the Moon's visibility window on each of the two candidate nights.
The first of these columns refers to the preceding night, from previous dusk to dawn, and the second to the recorded night, from dusk to the following dawn.
A dash (---) marks a night on which the Moon lay below the horizon throughout the night at the capital, so no visibility window exists and no midpoint can be given.

From Table~\ref{tab:events} we list some representative emendations that the monograph proposes.
For comparison, we consider another possibility: a guest star too faint to attract notice on its own could have drawn the observer's attention when the Moon occulted or closely passed it that night.
Because a planet lay nearby, or had recently moved through the same region, the observer may have recorded the transient as a lunar occultation of that planet.
The Moon shifts roughly 13° per day against the background stars, so such an alignment would not recur and the event would be recorded only once.
We offer this mechanism as one possible explanation, not an established identification.

\begin{itemize}
\item The monograph reads entries 7756 (1057 Jul 14) and 10786 (1284 Jan 13) as encounters with the Great Fire (\textit{dahuo}, Antares) of Xin rather than Mars.
Antares and Mars are similar in brightness and color.
The monograph explains a scribal chain in which ``fire'' was rewritten as ``Mars'' (\textit{huoxing}) and then as ``Yinghuo.''
Similarly, for entries 10868 (1295 Feb 23) and 11125 (1406 Feb 15), the monograph explains a scribal chain in which the star \textit{yuexing} was rewritten as ``Moon'' and then as \textit{taiyin}.

\item The monograph classifies entries 11995 (1736 May 08) and 12246 (1741 Jan 14) as computed rather than observed.
The occultation occurred while the Moon lay below the horizon, so it was invisible from China.
By the time the Moon rose, it had moved farther from the planet.
For 11995 the Moon rose briefly before dawn on 1736 May 08, and its position is given in the preceding-night column of Table~\ref{tab:events}.

\item The monograph reads entry 7171 (927 Jan 01) as a halo rather than an encroachment, since the Moon lay 17° from Venus, 7° from Jupiter, and 19° from Saturn, matching a halo radius of about 22°.

\item The monograph emends entry 3653 (488 Jul 03) from three cun to three chi.
One chi is about 1° and one cun about 0.1°, so three cun is about 0.3°.
The recomputation puts Jupiter 3.80° from the Moon, near three chi rather than three cun.

\item The monograph considers the dates of seventeen entries to be off by one day.
The event times fall outside the post-midnight hours described in Section~\ref{sec:computational_methods}, so the one-day offsets cannot be explained by the post-midnight dating convention.
These are entries 6749 (807 Mar 07), 6754 (816 Dec 25), 6779 (823 Feb 12), 6859 (835 Nov 02), 6869 (837 Mar 15), 7226 (944 Aug 01), 7474 (1277 Sep 11), 7711 (1011 Sep 25), 7716 (1020 Mar 10), 7721 (1021 Oct 15), 7726 (1038 Apr 19), 10945 (1308 Feb 16), 11120 (1379 Mar 19), 11132 (1445 Dec 20), 11137 (1449 Apr 24), 11142 (1454 Mar 04), and 11147 (1456 May 18).
Three entries carry additional caveats.
For entry 6869, the shifted date places the occultation on the morning of 837 Mar 14, a daytime event visible only in exceptionally clear weather.
Entry 7474 reverses the subject and object, recording that Mars occulted the Moon, and its description of a red sky remains unexplained.
For entry 11137, the Moon and Venus have a minimum angular separation of at least 2$^{\circ}$ on the shifted date, so the event does not qualify as an encroachment under the 2$^{\circ}$ criterion.

\item For fourteen entries the monograph notes that the record also appears in another source: 2709 (322 Jan 06), 2886 (359 Apr 20), 3007 (405 Jun 11), 3187 (434 May 09 and 434 May 20), 7681 (978 Aug 22 and 978 Sep 26), 7686 (988 Mar 11), 7691 (992 May 04), 7711 (1011 Sep 25), 7731 (1041 Nov 16), 11120 (1379 Mar 19), 11132 (1445 Dec 20), 11137 (1449 Apr 24), 11142 (1454 Mar 04), and 11147 (1456 May 18).
\end{itemize}

Although many entries can be attributed to copying or dating errors, some records are of particular interest and remain natural candidates for testing the guest-star hypothesis.
If a recorded planet was actually a transient, a remnant could lie near the candidate position under the guest-star hypothesis.
Modern searches for diffuse ejecta and compact objects near that position could therefore directly test this prediction.
A supernova, a classical nova, and a kilonova could each leave extended ejecta, producing a supernova remnant, a nova shell, or a kilonova remnant, respectively.
A core-collapse supernova or a binary neutron-star merger could also leave a neutron star, which may be observable as a pulsar or a source of continuous gravitational waves.
If future searches establish associations between remnants and these records, the corresponding records could become dated guest-star candidates, and their historical dates could anchor the remnant ages.

Another set of entries, however, contains observations that the guest-star hypothesis alone cannot fully reproduce, as the following cases demonstrate.
\begin{itemize}
\item Four entries carry dashes on both nights: entry 4368 (420 Dec 21), entry 5979 (886 May 08), entry 7691 (992 May 04), and entry 11120 (1379 Mar 19).
No occultation could therefore have been seen on those dates or on the nights preceding them.
Since the guest-star hypothesis cannot reproduce these records, a dating error is the more likely explanation.
For entry 7691, the monograph notes that the Moon and Mercury were only 1.4$^{\circ}$ apart on the morning of 992 May 04.
The record may therefore refer to this close approach rather than to an occultation of Venus.

\item Ten entries name a mansion that the Moon was not in on the recorded night: entry 2709 (322 Jan 06), entry 2886 (359 Apr 20), entry 2967 (400 Jul 23), entry 3007 (405 Jun 11), entry 4333 (412 Feb 18), entry 3187 (434 May 09 and 434 May 20), entry 4653 (529 Nov 19), entry 11167 (1586 Oct 09), entry 11172 (1607 Jun 27), and entry 12246 (1741 Jan 14).
If the recorded planet was the guest star, it must also have been near the named mansion.
The monograph supplies a corrected sexagenary day or month for nine of the ten entries, and each correction makes the recorded mansion correct.
Since the guest-star hypothesis cannot reproduce these mansion clauses, a dating error is the more likely explanation in these nine cases.

\item In entries 7731 (1041 Nov 16), 7746 (1056 Apr 01), 7751 (1056 Jun 22), and 10256 (1289 Aug 05), the monograph substitutes a different planet for the recorded one.
If a guest star had also been near the Moon on the relevant night, observers would be expected to describe multiple occulted bodies, but no such description appears.
These records are therefore more likely to reflect a misidentification of one planet as another.

\item For twelve entries a parallel source gives a different text or date, and the monograph accepts the parallel reading as correct and the recorded one as wrong: 1924 (303 Dec 05), 1944 (338 May 24 and 338 Aug 28), 2967 (400 Jul 23), 4333 (412 Feb 18), 6125 (712 Sep 21), 7696 (994 Mar 31), 7741 (1056 Jan 07 and 1056 Jan 10), 7756 (1057 Jul 14), 7771 (1209 Jul 06 and 1209 Jul 25), 11125 (1390 Dec 21 and 1406 Feb 15), 11157 (1478 Apr 08), and 11167 (1586 Oct 09).
A guest star would have appeared on the same night to all observers, so it cannot explain the disagreement; a copying error is the more likely explanation.
\end{itemize}

The remaining seventeen entries do not belong to any of the errata categories listed in the preceding paragraphs. 
They are 3328 (466 May 25), 3618 (483 Oct 21), 3690 (492 Jan 17), 3767 (494 Aug 09), 4480 (480 Mar 15), 4952 (498 Mar 21 and 498 May 01), 5046 (539 Apr 22), 5959 (839 Aug 28), 6844 (833 Jul 11), 6884 (839 Apr 02), 6919 (900 Dec 17), 7339 (1040 Mar 21), 7344 (1046 Apr 22), 7701 (1004 Sep 07), 7778 (1237 Aug 04), 10408 (1333 Oct 12), and 10418 (1334 Apr 07). 
Under the guest-star hypothesis, these records remain candidates, and the midpoint coordinates in Table~\ref{tab:events} may serve as candidate positions for remnant searches.

\section{Discussion}
\label{sec:discussion}

Liu Ci-Yuan's (2015) monograph systematically verifies thousands of entries in the Twenty-Four Histories and the Draft History of the Qing Dynasty against modern ephemerides.
Each entry quotes the original text, to which it adds verification analysis and textual criticism.
Its diagnoses attribute most failures to conventional causes: wrong dates, misidentification involving fixed stars or planets, and defects in transmission.
From the monograph we select the records of lunar occultations of planets, which form our catalog.
Such records allow us to recover what actually happened, because modern ephemerides can reconstruct the sky of the recorded night.
Because the monograph's diagnoses do not consider transient sources, we ask whether the recorded ``planet'' may instead have been a guest star.

We hypothesize that some guest stars were recorded only because the Moon occulted them.
Even a guest star too faint to attract notice on its own could have been recorded, because the Moon's occultation or close approach that night drew the observer's attention.
The Moon shifts roughly 13$^{\circ}$ per day, so the same alignment would not recur and the event would be recorded only once.
Guest stars such as novae and kilonovae can rival planets in naked-eye brightness, so ancient observers may have confused the two.
Planetary events may have been recorded as guest stars; conversely, guest stars may have been misidentified as planets.
The AD 1163 event illustrates the first direction: Chinese records describe a lunar encroachment (fan) on Saturn on 1163 Aug 10.
A Korean record of the same night instead reports a guest star, so those observers may have mistaken Saturn for a transient~\citep{2019MNRAS.490L..21L}.

This hypothesis is testable because the Moon's trajectory on the recorded night confines the transient to a narrow band.
Table~\ref{tab:events} gives the midpoint coordinates $(\alpha, \delta)$ of the Moon's visibility window on the two candidate nights.
We report two nights because the post-midnight convention leaves the recorded date ambiguous, as explained in Section~\ref{sec:computational_methods}.
The narrow band therefore defines a search region only a few degrees across, far narrower than the ten-odd degrees of uncertainty in textual asterism descriptions.
The AD 1163 benchmark confirms that the workflow reproduces a recorded event, with a minimum Moon--Saturn separation of 0.631$^{\circ}$ (Figure~\ref{fig:skyview_1163}).
Remnant searches near these candidate positions can therefore test the hypothesis, rather than merely suggest that a guest star appeared.

Among the 67 entries, the guest-star hypothesis alone cannot reproduce the records in four representative categories, as detailed in Section~\ref{sec:results}.
Four entries show no visibility window on either night, so the guest-star hypothesis cannot reproduce them and a dating error is more likely.
Ten entries name a mansion the Moon did not occupy on the recorded night, so the guest-star hypothesis cannot reproduce that mansion either.
For nine of them, the monograph's corrected date makes the recorded mansion correct, pointing again to a dating error as the more likely explanation.
The monograph substitutes a different planet in four entries, and a parallel source gives a different text or date in twelve, differences the guest-star hypothesis cannot explain.
These records therefore more likely reflect planetary misidentification or copying errors than a guest star.

The record alone cannot decide whether a failed entry reflects a guest star or an ordinary recording error.
The catalog therefore supplies candidate positions rather than identifications.
These limitations suggest several directions for future research.
We plan to cross-match the midpoint coordinates against known supernova remnant catalogs and to search the bands for diffuse ejecta such as nova shells and kilonova remnants.
A core-collapse supernova or a binary neutron-star merger could also leave a neutron star, so pulsar and continuous gravitational-wave searches offer a further test.
Recurrent novae offer another possible association, because a repeating transient could produce more than one record in separate outbursts.
If such associations are established, the records could become dated guest-star candidates, and their historical dates could anchor the remnant ages.

Even if those searches find no associated remnant, the null result would itself be informative.
The remnants may be too faint for current surveys, or the events may have been classical novae rather than supernovae.
The records may also reflect the copying, dating, or identification errors discussed above, in which case no transient was involved.
We would therefore still narrow the possible explanations, rather than simply refuting the guest-star hypothesis.

\section{Conclusions}
\label{sec:conclusions}

Liu Ci-Yuan's (2015) monograph provides our starting sample because its entry-by-entry textual criticism and checks against modern ephemerides cover the Twenty-Four Histories and the Draft History of the Qing Dynasty.
These checks supply entries already judged erroneous, so we narrowed the search space to 67 errata on lunar occultations of planets rather than rechecking every historical record.
The monograph attributes failures to wrong dates, identification errors involving fixed stars or planets, and defects in transmission, but it does not consider transient sources.
Guest star--planet confusion is nevertheless real and bidirectional: planets have been recorded as guest stars, whereas guest stars may have been recorded as planets.
Because guest stars and planets can appear similar in brightness and color to the unaided eye, we ask whether any of the planets recorded in these errata could instead have been guest stars.

We have publicly released the open-source Python package \texttt{occultsky} on GitHub so that researchers can reuse our reconstruction and testing workflow.
For each record, the package uses the recorded date and dynastic capital coordinates to construct a window anchored at solar noon and extending by $\pm 1$ day, covering the preceding and recorded nights as two candidates.
The package tests encroachment by checking whether the minimum Moon-planet angular separation falls below the 2$^{\circ}$ threshold.
For records that remain unconfirmed against modern ephemerides, it reconstructs the band swept by the lunar disk and projects it onto sky maps labeled with Chinese asterisms and star names.
The AD 1163 event serves as a benchmark because the workflow reproduces its recorded configuration in the corresponding reconstructed sky.

Our catalog comprises the 67 errata on lunar occultations of planets selected from Liu Ci-Yuan's (2015) monograph.
Table~\ref{tab:events} lists the apparent right ascension and declination $(\alpha, \delta)$ at the midpoint of the Moon's visibility window on two candidate nights for each record: the preceding night and the recorded night.
The two nights account for a recorded date that may be off by one day, as explained in Section~\ref{sec:computational_methods}.
The table uses \texttt{---} when the Moon remains below the horizon throughout a candidate night, so no visibility window exists.
These midpoint coordinates serve as proxies for the implied guest-star positions on each available candidate night.

The lunar trajectory confines a possible transient to a band only a few degrees across, giving remnant and compact-object searches a restricted region in which to test the guest-star hypothesis directly.
We provide candidate positions only and do not identify any specific remnant.
The guest-star hypothesis alone cannot explain four representative categories of entries; dating, transmission, or identification errors are more likely explanations for these records.
The other entries remain candidates for testing the hypothesis.
If future searches yield associations, the relevant records could supply dated guest-star candidates and age anchors for remnants.

Future work should test the hypothesis further by cross-matching remnant catalogs with the bands and searching for compact objects that could be associated with the historical events.
Even null results remain informative because they would constrain the range of possible interpretations of these records rather than simply refute the guest-star hypothesis.

\begin{acknowledgments}
We thank Liu Ci-Yuan for his monograph on the textual criticism of Chinese historical celestial records, which provided the starting sample for this work.
We thank Brandon Rhodes for developing and maintaining the open-source Skyfield package.
We thank the JPL Navigation and Ancillary Information Facility for providing the planetary ephemeris data.
This work was funded by the National Natural Science Foundation of China, Grant Numbers 12433006 and 12663007, the Guizhou Provincial Basic Research Program (Natural Science), Grant Number Qiankehejichu-[2024] qingnian 158, the Guizhou Provincial Fundamental Research Program, Grant Number QKHJCMS[2026]752, and the Guizhou University Talent Research Project (Natural Science), Grant Number GZUrenjihezi(2023)17.
During the preparation of this manuscript, the authors used AI-assisted tools for language editing and writing assistance only.
All scientific computations and figures were produced with \texttt{occultsky} and the other software described in this paper.
The authors reviewed and edited all content and take full responsibility for the content of this publication.
\end{acknowledgments}

\software{Skyfield~\citep{2019ascl.soft07024R}, Stellarium~\citep{2021arXiv210401019Z}, occultsky~\citep{Liu2026occultsky}}

\bibliography{references}{}

\begin{thebibliography}{}
\expandafter\ifx\csname natexlab\endcsname\relax\def\natexlab#1{#1}\fi
\providecommand{\url}[1]{\href{#1}{#1}}
\providecommand{\dodoi}[1]{doi:~\href{http://doi.org/#1}{\nolinkurl{#1}}}
\providecommand{\doeprint}[1]{\href{http://ascl.net/#1}{\nolinkurl{http://ascl.net/#1}}}
\providecommand{\doarXiv}[1]{\href{https://arxiv.org/abs/#1}{\nolinkurl{https://arxiv.org/abs/#1}}}

\bibitem[{S.-H. {Ahn}(2020){Ahn}}]{2020JKAS...53..125A}
{Ahn}, S.-H. 2020, \bibinfo{title}{{Observing Sites for the Central Solar
  Eclipses in Ancient Chinese History},} Journal of Korean Astronomical
  Society, 53, 125, \dodoi{10.5303/JKAS.2020.53.6.125}

\bibitem[{M.~R. {Calabretta} {\&} E.~W. {Greisen}(2002){Calabretta} \&
  {Greisen}}]{2002A&A...395.1077C}
{Calabretta}, M.~R., \& {Greisen}, E.~W. 2002, \bibinfo{title}{{Representations
  of celestial coordinates in FITS},} \aap, 395, 1077,
  \dodoi{10.1051/0004-6361:20021327}

\bibitem[{J.~J. {Hester}(2008){Hester}}]{2008ARA&A..46..127H}
{Hester}, J.~J. 2008, \bibinfo{title}{{The Crab Nebula : an astrophysical
  chimera.},} \araa, 46, 127, \dodoi{10.1146/annurev.astro.45.051806.110608}

\bibitem[{J.~L. {Hilton} {et~al.}(1988){Hilton}, {Seidelmann}, \&
  {Liu}}]{1988AJ.....96.1482H}
{Hilton}, J.~L., {Seidelmann}, P.~K., \& {Liu}, C. 1988,
  \bibinfo{title}{{Analysis of ancient Chinese records of occultations between
  planets and stars},} \aj, 96, 1482, \dodoi{10.1086/114900}

\bibitem[{J.~L. {Hilton} {et~al.}(1992){Hilton}, {Seidelmann}, \&
  {Liu}}]{1992AJ....104.2250H}
{Hilton}, J.~L., {Seidelmann}, P.~K., \& {Liu}, C. 1992, \bibinfo{title}{{An
  Examination of the Change in the Earth's Rotation Rate From Ancient Chinese
  Observations of Lunar Occultations of the Planets},} \aj, 104, 2250,
  \dodoi{10.1086/116401}

\bibitem[{P.~Y. {Ho}(1962){Ho}}]{1962VA......5..127H}
{Ho}, P.~Y. 1962, \bibinfo{title}{{Ancient and mediaeval observations of comets
  and novae in Chinese sources},} Vistas in Astronomy, 5, 127,
  \dodoi{10.1016/0083-6656(62)90007-7}

\bibitem[{T. {Kiang}(1981){Kiang}}]{1981ChA&A...5..111K}
{Kiang}, T. 1981, \bibinfo{title}{{On the date used in Chinese historical
  annals when recording observations made during the latter half of the
  night},} \caa, 5, 111, \dodoi{10.1016/0275-1062(81)90082-5}

\bibitem[{C.-Y. {Liu}(1988){Liu}}]{1988AcApS...8..147L}
{Liu}, C.-Y. 1988, \bibinfo{title}{{The secular variation of the earth's
  rotation obtained from 171 Chinese records of lunar occultation before A.D.
  600},} Acta Astrophysica Sinica, 8, 147

\bibitem[{C.~Y. {Liu} {\&} L.~P. {Ma}(2018){Liu} \& {Ma}}]{2018AcASn..59...28L}
{Liu}, C.~Y., \& {Ma}, L.~P. 2018, \bibinfo{title}{{Statistical Analysis on the
  Astronomical Records of Ming Shilu},} Acta Astronomica Sinica, 59, 28

\bibitem[{Y. {Liu}(2026){Liu}}]{Liu2026occultsky}
{Liu}, Y. 2026, {occultsky}, v0.1.0 Zenodo, \dodoi{10.5281/zenodo.24297100}

\bibitem[{Y. {Liu} {et~al.}(2019){Liu}, {Zou}, {Jiang}, {Gao}, {Ma}, \&
  {Liao}}]{2019MNRAS.490L..21L}
{Liu}, Y., {Zou}, Y.-C., {Jiang}, B., {et~al.} 2019, \bibinfo{title}{{G4.8+6.2,
  a possible kilonova remnant?},} \mnras, 490, L21,
  \dodoi{10.1093/mnrasl/slz141}

\bibitem[{A. {Mokhtarzadeh}(2026){Mokhtarzadeh}}]{2026AAS...24712503M}
{Mokhtarzadeh}, A. 2026, \bibinfo{title}{{Bridging the Skies: Global
  Constellations Through Science and Story},} in American Astronomical Society
  Meeting Abstracts, Vol. 247, American Astronomical Society Meeting Abstracts,
  125.03

\bibitem[{D.~L. {Neuh{\"a}user} {et~al.}(2025){Neuh{\"a}user}, {Neuh{\"a}user},
  {Hambaryan}, {Chapman}, \& {Della Valle}}]{2025A&A...704A.251N}
{Neuh{\"a}user}, D.~L., {Neuh{\"a}user}, R., {Hambaryan}, V., {Chapman}, J., \&
  {Della Valle}, M. 2025, \bibinfo{title}{{Challenging historical novae: AT Cnc
  (1645), Te-11 (483), and M22 (BC 48) revisited},} \aap, 704, A251,
  \dodoi{10.1051/0004-6361/202554643}

\bibitem[{B. {Rhodes}(2019){Rhodes}}]{2019ascl.soft07024R}
{Rhodes}, B. 2019, {Skyfield: High precision research-grade positions for
  planets and Earth satellites generator},, Astrophysics Source Code Library,
  record ascl:1907.024 \doeprint{1907.024}

\bibitem[{F.~R. {Stephenson}(2002){Stephenson}}]{2002HiA....12..317S}
{Stephenson}, F.~R. 2002, \bibinfo{title}{{East Asian astronomical records},}
  Highlights of Astronomy, 12, 317

\bibitem[{F.~R. {Stephenson} {\&} L.~V. {Morrison}(1995){Stephenson} \&
  {Morrison}}]{1995RSPTA.351..165S}
{Stephenson}, F.~R., \& {Morrison}, L.~V. 1995, \bibinfo{title}{{Long-Term
  Fluctuations in the Earth's Rotation: 700 BC to AD 1990},} Philosophical
  Transactions of the Royal Society of London Series A, 351, 165,
  \dodoi{10.1098/rsta.1995.0028}

\bibitem[{F.~R. {Stephenson} {\&} K.~K.~C. {Yau}(1985){Stephenson} \&
  {Yau}}]{1985JBIS...38..195S}
{Stephenson}, F.~R., \& {Yau}, K.~K.~C. 1985, \bibinfo{title}{{Far Eastern
  observations of Halley's comet: 240 BC to AD 1368.},} Journal of the British
  Interplanetary Society, 38, 195

\bibitem[{C.~G.~R. {Wallis} {et~al.}(2022){Wallis}, {Price}, {McEwen},
  {Kitching}, {Leistedt}, \& {Plouviez}}]{2022MNRAS.509.4480W}
{Wallis}, C. G.~R., {Price}, M.~A., {McEwen}, J.~D., {et~al.} 2022,
  \bibinfo{title}{{Mapping dark matter on the celestial sphere with weak
  gravitational lensing},} \mnras, 509, 4480, \dodoi{10.1093/mnras/stab3235}

\bibitem[{D.~M. {Willis} {\&} F.~R. {Stephenson}(2001){Willis} \&
  {Stephenson}}]{2001AnGeo..19..289W}
{Willis}, D.~M., \& {Stephenson}, F.~R. 2001, \bibinfo{title}{{Solar and
  auroral evidence for an intense recurrent geomagnetic storm during December
  in AD 1128},} Annales Geophysicae, 19, 289, \dodoi{10.5194/angeo-19-289-2001}

\bibitem[{S.-X. {Wu} {\&} C.-Y. {Li{\"u}}(1993){Wu} \&
  {Li{\"u}}}]{1993ChA&A..17..337W}
{Wu}, S.-X., \& {Li{\"u}}, C.-Y. 1993, \bibinfo{title}{{Secular variation of
  Earth's rotation from ancient Chinese records of lunar occultations and
  approaches},} \caa, 17, 337, \dodoi{10.1016/0275-1062(93)90101-T}

\bibitem[{Z.-Z. {Xi}(1955{\natexlab{a}}){Xi}}]{1955AcASn...3..183X}
{Xi}, Z.-Z. 1955{\natexlab{a}}, \bibinfo{title}{{A New Catalogue of Ancient
  Novae},} Acta Astronomica Sinica, 3, 183

\bibitem[{Z.-Z. {Xi} {et~al.}(1966){Xi}, {Po}, \& {Yang}}]{1966Sci...154..597X}
{Xi}, Z.-Z., {Po}, S.-J., \& {Yang}, K.~S. 1966, \bibinfo{title}{{Ancient
  Oriental Records of Novae and Supernovae},} Science, 154, 597,
  \dodoi{10.1126/science.154.3749.597}

\bibitem[{G. {Zotti} {et~al.}(2021){Zotti}, {Hoffmann}, {Wolf}, {Ch{\'e}reau},
  \& {Ch{\'e}reau}}]{2021arXiv210401019Z}
{Zotti}, G., {Hoffmann}, S.~M., {Wolf}, A., {Ch{\'e}reau}, F., \&
  {Ch{\'e}reau}, G. 2021, \bibinfo{title}{{The Simulated Sky: Stellarium for
  Cultural Astronomy Research},} arXiv e-prints, arXiv:2104.01019,
  \dodoi{10.48550/arXiv.2104.01019}

\end{thebibliography}
\bibliographystyle{aasjournalv7.1}

\end{document}